\documentclass[prd,twocolumn,showpacs,eqsecnum]{revtex4}
\usepackage{graphicx}

\begin{document}

\title{Neutral-Current Atmospheric Neutrino Flux Measurement
Using Neutrino-Proton Elastic Scattering in Super-Kamiokande}

\author{
\mbox{John F. Beacom$^{1}$} and
\mbox{Sergio Palomares-Ruiz$^{1, 2}$}}

\affiliation{
\mbox{$^1$  NASA/Fermilab Astrophysics Center, Fermi National 
Accelerator Laboratory, Batavia, Illinois 60510-0500, USA}
\mbox{$^2$  Departamento de F\'{\i}sica Te\'orica, Universidad de 
Valencia, 46100 Burjassot, Valencia, Spain}
\\
{\tt beacom@fnal.gov},
{\tt Sergio.Palomares@uv.es}}

\date{January 9, 2003}

\begin{abstract}
Recent results show that atmospheric $\nu_\mu$ oscillate with $\delta
m^2 \simeq 3 \times 10^{-3}$ eV$^2$ and $\sin^2{2\theta_{atm}} \simeq
1$, and that conversion into $\nu_e$ is strongly disfavored.  The
Super-Kamiokande (SK) collaboration, using a combination of three
techniques, reports that their data favor $\nu_\mu \rightarrow
\nu_\tau$ over $\nu_\mu \rightarrow \nu_{sterile}$.  This distinction
is extremely important for both four-neutrino models and cosmology.
We propose that neutrino-proton elastic scattering ($\nu + p
\rightarrow \nu + p$) in water \v{C}erenkov detectors can also
distinguish between active and sterile oscillations.  This was not
previously recognized as a useful channel since only about 2\% of
struck protons are above the \v{C}erenkov threshold.  Nevertheless, in
the present SK data there should be about 40 identifiable events.  We
show that these events have unique particle identification
characteristics, point in the direction of the incoming neutrinos, and
correspond to a narrow range of neutrino energies ($1-3$ GeV,
oscillating near the horizon).  This channel will be particularly
important in Hyper-Kamiokande, with $\sim 40$ times higher rate.  Our
results have other important applications.  First, for a similarly
small fraction of atmospheric neutrino quasielastic events, the proton
is relativistic.  This uniquely selects $\nu_\mu$ (not
$\bar{\nu}_\mu$) events, useful for understanding matter effects, and
allows determination of the neutrino energy and direction, useful for
the $L/E$ dependence of oscillations.  Second, using accelerator
neutrinos, both elastic and quasielastic events with relativistic
protons can be seen in the K2K 1-kton near detector and MiniBooNE.
\end{abstract}

\pacs{13.15.+g, 25.30.Pt, 29.40.Ka 
%Neutrino interactions, Nuclear neutrino scattering, Cherenkov detectors       
\hspace{0.7cm} FERMILAB-Pub-03/003-A, FTUV-03-0109, IFIC/03-01}

\maketitle

%%%%%%%%%%%%%%%%%%%%%%%%%%%%%%%%%%%%%%%%%%%%%%%%%%%%%%%%%%%%%%%%%%%%%%%%%%
%                           Introduction                                 %
%%%%%%%%%%%%%%%%%%%%%%%%%%%%%%%%%%%%%%%%%%%%%%%%%%%%%%%%%%%%%%%%%%%%%%%%%%

\section{Introduction}

High-statistics atmospheric neutrino data from Super-Kamiokande (SK)
show $\nu_\mu$ vacuum oscillation disappearance with $\delta m^2
\simeq 3 \times 10^{-3}$ eV$^2$ and $\sin^2{2\theta_{atm}} \simeq
1$~\cite{SKatm}.  Both atmospheric~\cite{SKatm} and
reactor~\cite{reactor} data strongly disfavor $\nu_\mu \rightarrow
\nu_e$ oscillations, so one of the crucial remaining questions is
whether the oscillations are of $\nu_\mu \rightarrow \nu_\tau$ or
$\nu_\mu \rightarrow \nu_{sterile}$, or a mixture.  A definite answer
would have important implications for four-neutrino mixing models
designed to accommodate~\cite{fournulab} the LSND signal~\cite{lsnd}
and also for the role of neutrinos in cosmology~\cite{fournucosmo}.

A variety of techniques have been proposed to distinguish between
atmospheric $\nu_\mu \rightarrow \nu_\tau$ and $\nu_\mu \rightarrow
\nu_{sterile}$
oscillations~\cite{sterilePi,sterileME,sterileMR,sterileTau,sterileDK}.
The SK collaboration reports that their data favor pure active
oscillations $\nu_\mu \rightarrow \nu_\tau$ over pure sterile
oscillations $\nu_\mu \rightarrow \nu_{sterile}$ at better than the
99\% CL~\cite{SKNC}, though an appreciable sterile admixture remains
possible.  This claim is based on a combination of three techniques:
(i) matter effects on the partially-contained and throughgoing-muon
samples, (ii) the contribution of inelastic neutral-current events
with pions to the fully-contained multi-ring events sample, and (iii)
statistically-identified tau lepton decays.  Individually, these
techniques do not offer strong evidence, and the latter two are made
difficult by large backgrounds.

We propose a new technique for active-sterile discrimination:
neutrino-proton elastic scattering,
\begin{equation}
\nu + p \rightarrow \nu + p\,,
\end{equation}
from both free and bound protons, where the struck proton is above the
\v{C}erenkov threshold (see also a related detection application at
low energies in Ref.~\cite{elastic}).  This is a neutral-current (NC)
channel, and so directly measures the total active neutrino flux.  The
number of charged-current (CC) atmospheric neutrino events in SK is
$\sim 10^4$, and the ratio of cross sections $\sigma_{NC}/\sigma_{CC}
\simeq 0.2$ for both neutrinos and antineutrinos in this energy
range~\cite{Ahrens}, so we expect $\sim 10^3$ NC events.  It has
always been assumed that protons above the \v{C}erenkov threshold are
too small in number and too difficult to identify.  We show that the
number of events is reasonable and that the protons are identifiable.
The protons are directional, and an up-down ratio discriminates
between active and sterile oscillations since upgoing $\nu_\tau$ are
fully detectable while upgoing $\nu_{sterile}$ are invisible.

This detection channel for atmospheric neutrinos in SK was mentioned
by Vissani and Smirnov, but was dismissed as having too few events
above the \v{C}erenkov threshold~\cite{sterilePi}.  Some estimates for
the event rate were mentioned in SK Ph.D. theses~\cite{SKtheses}.  In
those studies, all of the protons were classified by the Monte Carlo
as either $e$-like or $\mu$-like, and were assumed to be buried by the
much more numerous relativistic electrons and muons from CC channels.
It is unknown what fraction of relativistic protons failed general
cuts designed for electrons and muons, and hence were not classified
at all.

In what follows, we develop the case that atmospheric neutrino-proton
elastic scattering events above the \v{C}erenkov threshold should be
in the present SK data set, that they can be identified, and that they
will be very useful in studying neutrino oscillations.  Finally, we
identify some other applications of our results.

%%%%%%%%%%%%%%%%%%%%%%%%%%%%%%%%%%%%%%%%%%%%%%%%%%%%%%%%%%%%%%%%%%%%%%%%%%%%
%                             Cross section                                %
%%%%%%%%%%%%%%%%%%%%%%%%%%%%%%%%%%%%%%%%%%%%%%%%%%%%%%%%%%%%%%%%%%%%%%%%%%%%

\section{Cross Section for $\nu + p \rightarrow \nu + p$}

%%%%%%%%%%%%%%%%%%%%%%%%%%%%% Free protons %%%%%%%%%%%%%%%%%%%%%%%%%%%%%%%%%

\subsection{Free Proton Targets}

%%%%%%%%%%%%%%
\begin{figure}
\includegraphics[width=3.25in]{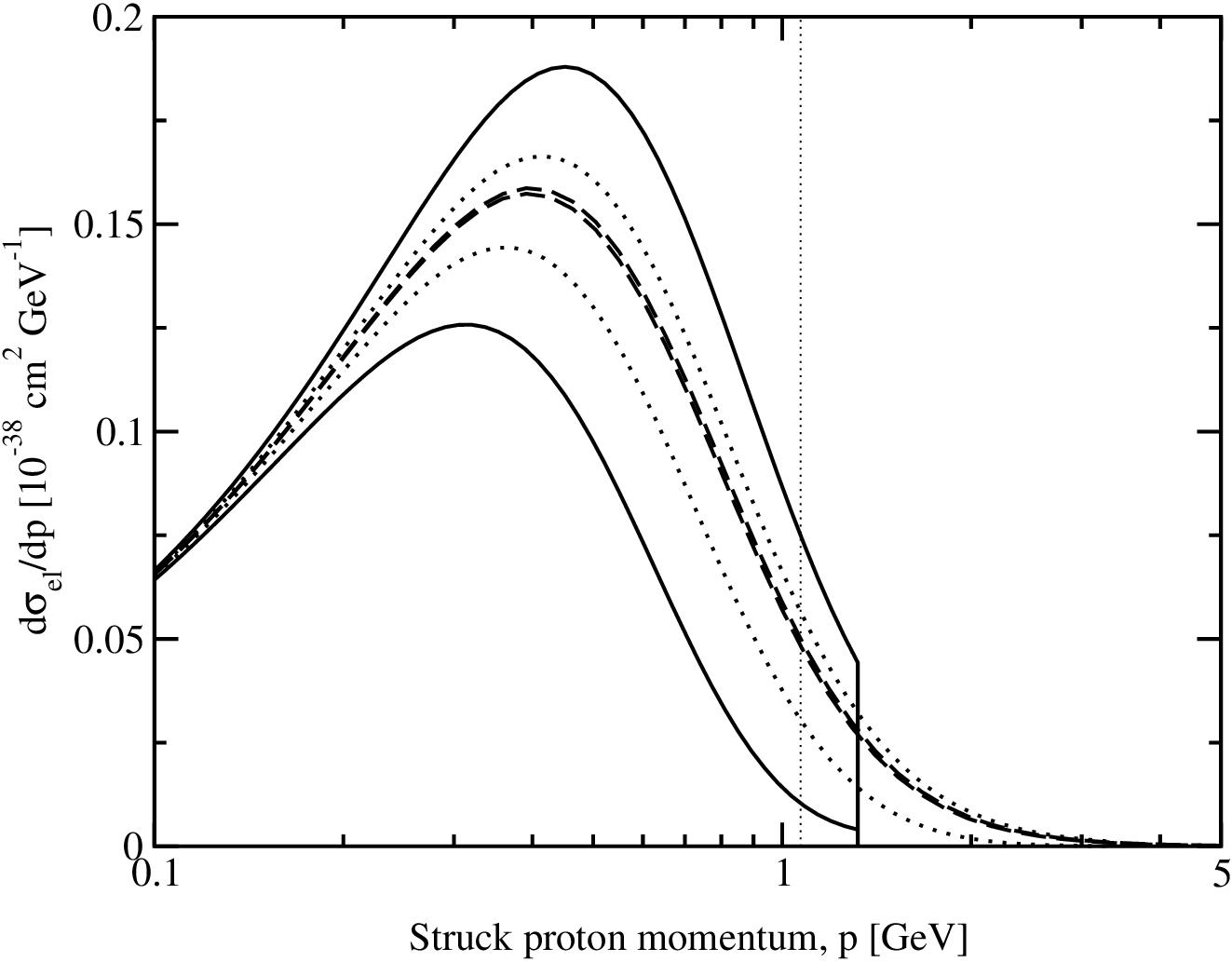}
\caption{\label{dsdp} Differential $\nu + p \rightarrow \nu + p$ cross
section as a function of proton momentum (note log scale used for
display).  Upper lines correspond to neutrinos and lower lines to
antineutrinos, at $E_\nu = 1$ GeV (solid), 3 GeV (dotted), and 50 GeV
(dashed), terminated at the maximum allowed proton momentum (only
visible in the 1 GeV case).  The \v{C}erenkov threshold in water at $p
= 1.07$ GeV is shown with a thin dotted line.}
\end{figure} 
%%%%%%%%%%%%%%

The neutrino-proton elastic scattering cross section is an important
prediction of the Standard Model that has been confirmed at GeV
energies, e.g., in the E734 experiment~\cite{Ahrens}.  The
differential cross section on free proton targets in terms of the
struck proton momentum $p$ (and corresponding mass $M_p$, kinetic
energy $T_p$, and total energy $E_p$) and neutrino energy $E_\nu$
is given by~\cite{cs}
\begin{equation}
\label{crosssect}
\frac{d\sigma_{el}}{dp} = \frac{G_F^2 M_p^3 p}{4 \pi E_{\nu}^2 E_p}
%\sqrt{1 - \left(\frac{M_p}{T_p + M_p}\right)^2}
\left[A \mp B \frac{(s-u)}{M_p^2} + C \frac{(s-u)^2}{M_p^4}\right]\,.
\end{equation}
The minus (plus) is for neutrinos (antineutrinos), and
\begin{eqnarray}
\label{abc}
A & = & 
4 \tau \left[G_A^2 (1 + \tau) - F_{1}^2 (1 - \tau) \right. \nonumber \\
& + & \left. F_2^2 (1 - \tau) \tau + 4 F_1 F_2 \tau \right]\\
B & = & 4 \tau G_A (F_1 + F_2) \\
C & = & \frac{1}{4} \left(G_A^2 + F_1^2 + F_2^2 \tau\right)\,,
\end{eqnarray}
where $s - u = 4 M_p \left(E_\nu - M_p \tau \right)$, $\tau = Q^2/4
M_p^2$, and $Q^2 = -q^2 = 2 M_p T_p$.  The differential cross section
is shown in Fig.~\ref{dsdp}.  Almost all of the struck protons are
below the \v{C}erenkov threshold; the subject of this paper are those
few above it.  Since this is a neutral-current channel, all active
flavors of neutrinos contribute equally (though the antineutrino cross
section is smaller at the relevant $E_\nu$).  We consider backgrounds
to the detection of protons in the elastic channel in detail below.

\begin{figure}
\includegraphics[width=3.25in]{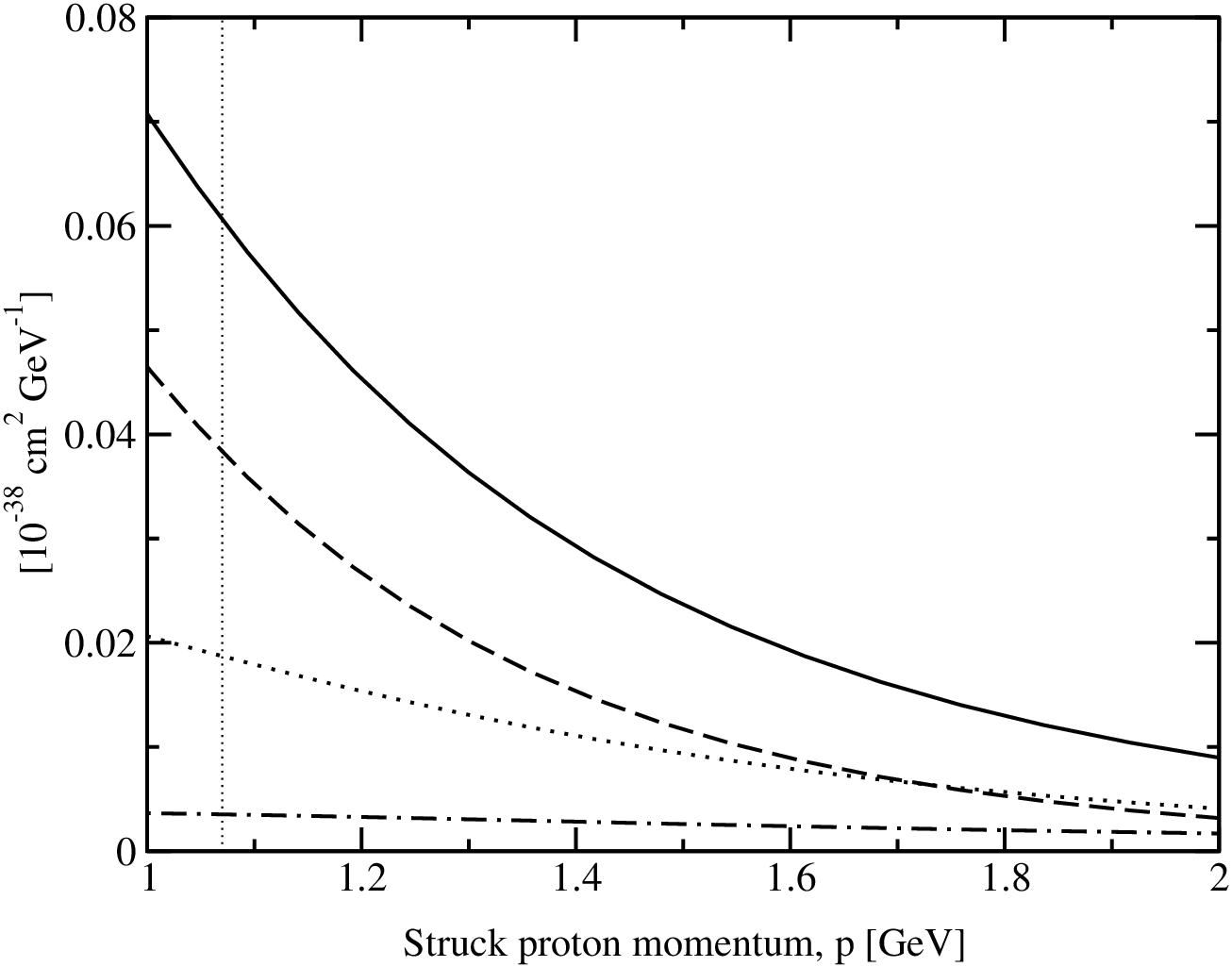}
\caption{\label{terms} The components of the differential $\nu + p
\rightarrow \nu + p$ cross section: the sum (solid line), $A$ term
(dot-dashed line), $B$ term (dotted line) and $C$ term (dashed line)
for $E_\nu = 2$ GeV, as a function of the recoil proton momentum $p$.
For antineutrinos, the $B$ term contributes with the opposite sign,
and so the differential cross section is much smaller.  Note that only
$p > 1$ GeV is shown, in contrast to Fig.~\ref{dsdp}.}
\end{figure}

The most important proton recoil momenta lie between about 1 and 2
GeV.  The lower limit is determined by the \v{C}erenkov threshold in
water at 1.07 GeV, and the upper limit by the falling differential
cross section (and neutrino spectrum).  In Fig.~\ref{terms}, we show
the separate terms of the differential cross section for $E_\nu = 2$
GeV (other relevant energies give similar results).  The most
important terms in Eq.~(\ref{crosssect}) are the $B$ and $C$ terms,
which are comparable.  For neutrinos, they add constructively, and for
antineutrinos, they add destructively.  The suppression of the
differential cross section at large proton momenta is caused by the
decrease of the form factors at large $Q^2$.

%%%%%%%%%%%%%%
%%%%%%%%%%%%%%

The vector form factors $F_1$ and $F_2$ that couple the exchanged
$Z^\circ$ to the proton have been well measured in electron-nucleon
scattering, and are
\begin{eqnarray}
\label{f}
F_1 & = & (1 - 2 \sin^2{\theta_W}) (G_V^3 - F_V^3) \nonumber \\
& - & \frac{2}{3} \sin^2{\theta_W} (G_V^0 - F_V^0) \\
F_2 & = & 
(1 - 2 \sin^2{\theta_W}) F_V^3 - \frac{2}{3}\sin^2{\theta_W} F_V^0\,,
\end{eqnarray}
where $\sin^2{\theta_W} = 0.231$.  The remaining form factors have a
dipole form (vector mass $M_V = 0.84$ GeV), and are
\begin{eqnarray}
\label{formfact}
G_V^3 & = & \frac{1}{2} \frac{1 + \kappa_p - \kappa_n}{(1 + Q^2/M_V^2)^2}\\
G_V^0 & = & \frac{3}{2} \frac{1 + \kappa_p + \kappa_n}{(1 + Q^2/M_V^2)^2}\\
F_V^3 & = & \frac{1}{2} 
\frac{\kappa_p - \kappa_n}{(1 + \tau) (1 + Q^2/M_V^2)^2}\\
F_V^0 & = & \frac{3}{2} 
\frac{\kappa_p + \kappa_n}{(1 + \tau) (1 + Q^2/M_V^2)^2}\,,
\end{eqnarray}
where $\kappa_p = 1.793$ and $\kappa_n = - 1.913$ are the proton and
neutron anomalous magnetic moments.

The axial form factor is assumed to have a dipole form (axial mass
$M_A = 1.03$ GeV), given by
\begin{equation}
\label{ga}
G_A = - \frac{1}{2} \frac{1.267}{(1 + Q^2/M_A^2)^2}\,.
\end{equation}
The main uncertainty in the differential cross section,
Eq.~(\ref{crosssect}), is caused by uncertainties in $G_A$.  While
neutrino scattering experiments suggest $M_A = 1.03$ GeV, charged-pion
electroproduction data~\cite{axial2} suggest $M_A = 1.08$ GeV; all of
the data are reviewed in Ref.~\cite{axial1}.  The larger value of
$M_A$ would increase the differential cross section by about 10\% in
the interval we are interested in.  There can also be strange sea
quark contributions to all of the form factors, especially the spin
contribution $\Delta s$ that modifies the axial form
factor~\cite{strange}, possibly increasing the differential cross
section by about 10\%.  However, the effects of changing $\Delta s$
and $M_A$ are correlated.  We do not assume enhancements to the cross
section from either a larger $M_A$ or a $\Delta s$ contribution.

%%%%%%%%%%%%%%%%%%%%%%%%% Nuclear Effects %%%%%%%%%%%%%%%%%%%%%%%%%%%

\subsection{Nuclear Effects}

We have so far only considered free protons, which are 2 out of 10
targets in water.  Bound protons have nonzero initial momenta (Fermi
motion), and the struck protons cannot make transitions to
already-filled states at low energies (Pauli blocking).  In typical
Fermi-gas models ($p_F \simeq 220$ MeV) for neutrino interactions at a
few GeV, these effects reduce the total cross section by about
20\%~\cite{Gaisser}.  However, these effects can be neglected when the
struck proton is above the \v{C}erenkov threshold.  In this limit, the
struck proton is ejected from the nucleus and the momentum transfer
greatly exceeds the initial momentum.  This may be seen from the
differential cross section results (for the CC channel) in
Refs.~\cite{Gaisser,Co}.

The struck proton may reinteract as it leaves the nucleus~\cite{Co};
at the relevant momenta, the interaction probability is about 1/2,
corresponding mostly to forward elastic collisions~\cite{Walter}.
There are neutrino interaction codes~\cite{nucodes} that take nuclear
reinteractions into account, but we do not.  The average momentum loss
for protons bound in oxygen in quasielastic scattering using the K2K
neutrino beam is only $\simeq 90$ MeV~\cite{Walter}; it is reasonable
to assume a similar average value for elastic atmospheric neutrino
events.  Since the spectrum $dN/dp$ is so steep (see below), taking
this into account could reduce the number of events above the
\v{C}erenkov threshold by about 20\%.  We neglect this effect because
we are also neglecting the fact that there would be some compensation
to this loss from $\nu + n \rightarrow \nu + n$ (about 1.5 times
larger than for protons) followed by a $n + p \rightarrow p + n$
nuclear reinteraction that transfers most of the momentum to the
proton.  A full detector Monte Carlo will be needed to model these
effects, along with the details of the \v{C}erenkov particle
identification and backgrounds.

%%%%%%%%%%%%%%%%%%%%%%%%  Angular distribution %%%%%%%%%%%%%%%%%%%%%%%%%%

%%%%%%%%%%%%%%
\begin{figure}
\includegraphics[width=3.25in]{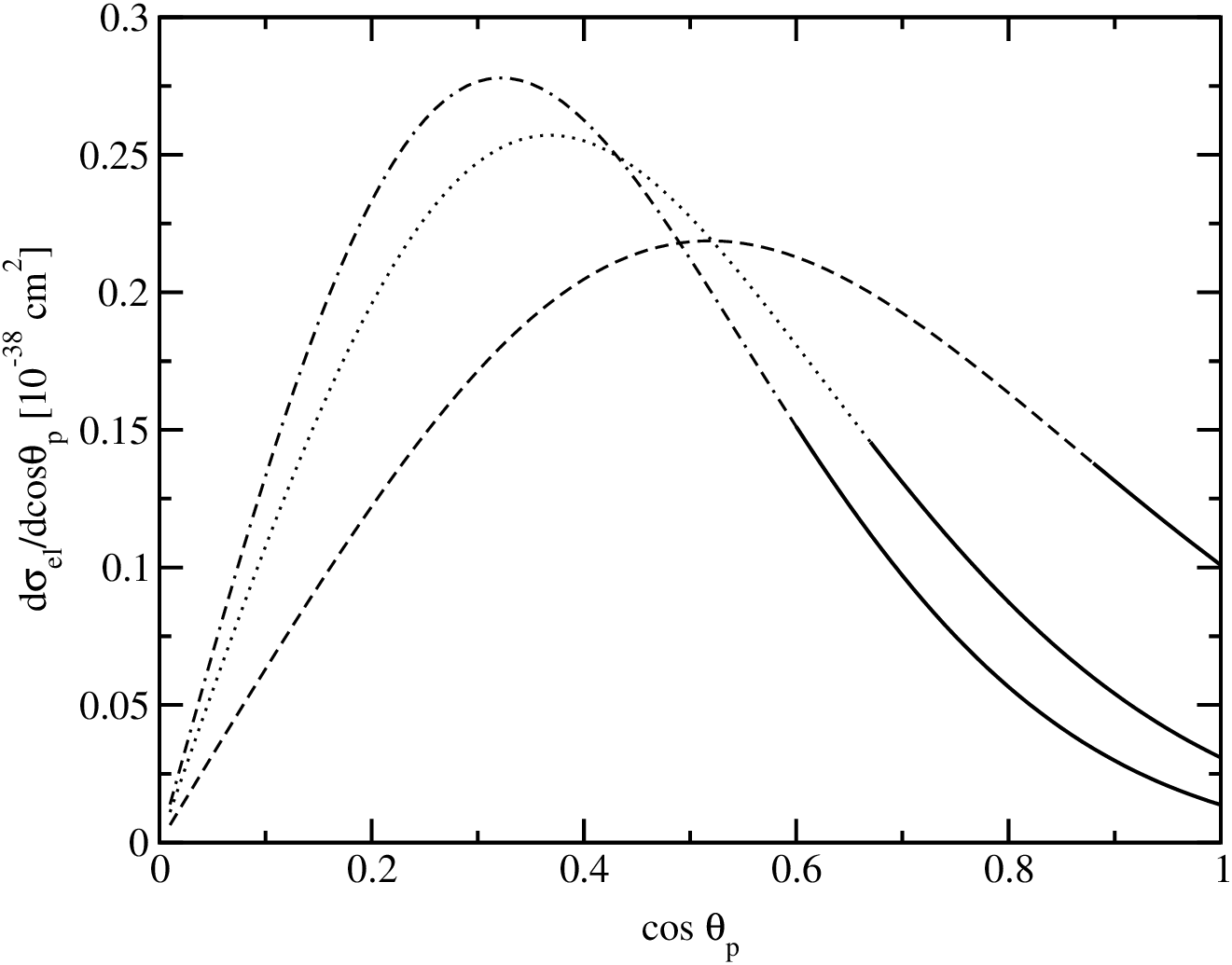}
\caption{\label{dsdth} The differential cross section
$d\sigma_{el}/d\cos{\theta_p}$ ($\theta_p$ is the angle of the struck
proton with respect to the incoming neutrino direction) for different
neutrino energies, $E_\nu$ = 1 (dashed line), 2 (dotted line), 3 GeV
(dot-dashed line).  The solid-line segment of each curve indicates
protons that are above the \v{C}erenkov threshold.  Since we are
neglecting proton reinteractions in the nucleus, only the forward
hemisphere is shown.}
\end{figure}
%%%%%%%%%%%%%%

\subsection{Angular Distribution}

In order to discriminate between active and sterile oscillations, the
struck protons must be directional.  The angles of the final particles
relative to the initial neutrino direction are given by
\begin{eqnarray}
\label{variables}
\cos{\theta_p} & = &
\frac{E_\nu + M_p}{E_\nu} \sqrt{\frac{T_p}{T_p + 2 M_p}}\\
\cos{\theta_\nu} & = &  1 - \frac{M_p T_p}{E_\nu (E_\nu - T_p)}\,.
\end{eqnarray}
The maximum proton momentum is obtained when the neutrino reverses its
direction and the proton goes forward.  The most important neutrino
energies for this channel are $1-3$ GeV, so for $\delta m^2 \simeq 3
\times 10^{-3}$ eV$^2$, the oscillation length corresponds to the
direction of the horizon.  Thus downgoing $\nu_\mu$ have not
oscillated yet, and upgoing $\nu_\mu$ have oscillated into either
$\nu_\tau$ (same rate as downgoing) or $\nu_{sterile}$ (a reduced
rate).  As shown in Fig.~\ref{dsdth}, most protons emerge at rather
large angles relative to the neutrino direction.  However, these are
the same majority of protons that are below the \v{C}erenkov
threshold.  The relevant protons above the \v{C}erenkov threshold are
in fact quite forward, but not perfectly so (we show below that the
lowest neutrino energies are the most relevant).  Compared to the
intrinsic angular variation, angular deflections from nuclear
reinteractions~\cite{Walter} can almost always be ignored.

%%%%%%%%%%%%%%%%%%%%%%%%%%%%%%%%%%%%%%%%%%%%%%%%%%%%%%%%%%%%%%%%%%%%%%%%%%
%                         Proton recoil spectrum                         %
%%%%%%%%%%%%%%%%%%%%%%%%%%%%%%%%%%%%%%%%%%%%%%%%%%%%%%%%%%%%%%%%%%%%%%%%%%

\section{Proton Recoil Spectrum}

%%%%%%%%%%%%%%%%%%%%%%%%%% No-Oscillation Prediction %%%%%%%%%%%%%%%%%%%%%

\subsection{No-Oscillation Prediction}

%%%%%%%%%%%%%%
\begin{figure}
\includegraphics[width=3.25in]{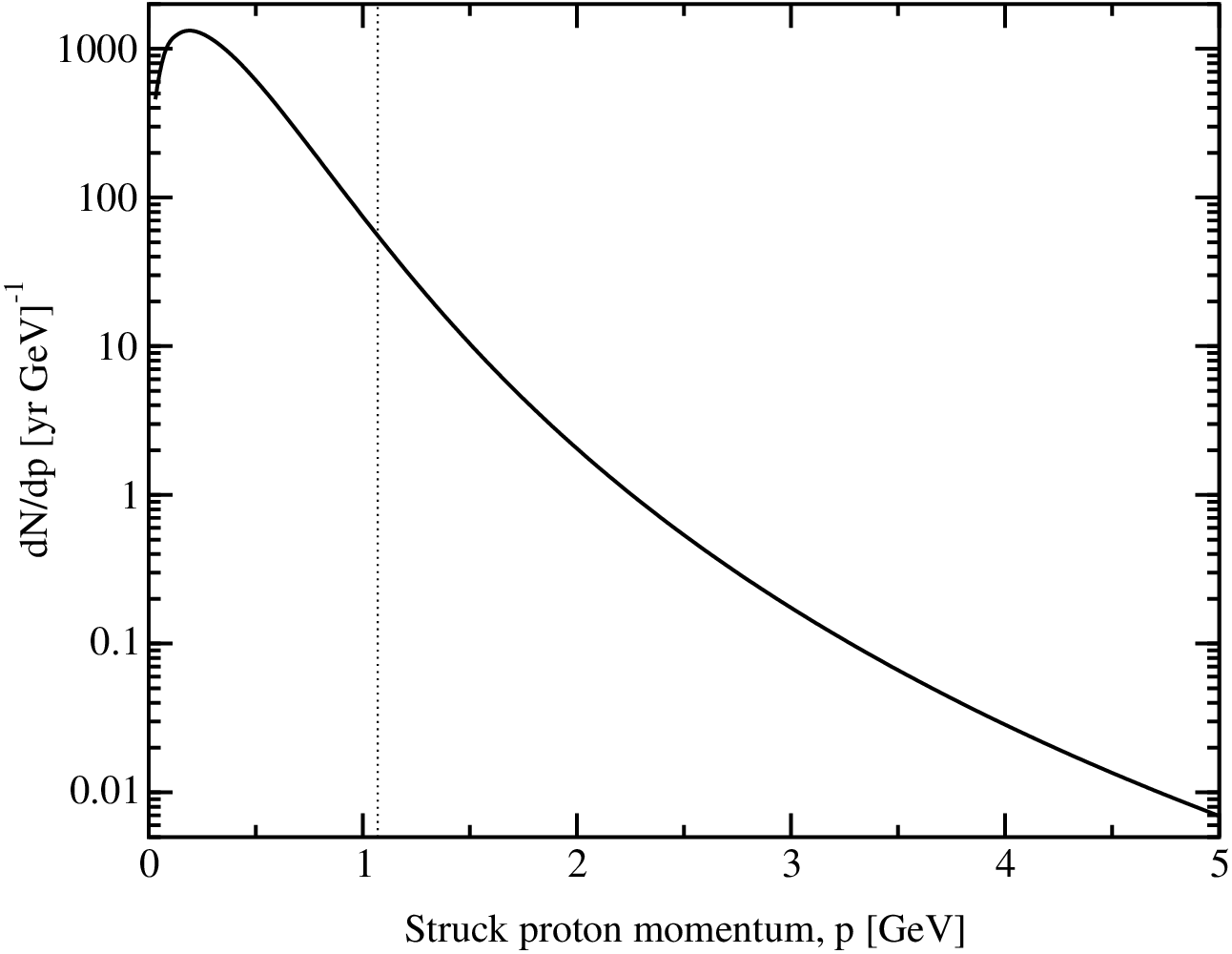}
\caption{\label{dNdpI} Proton spectrum, dN/dp (yr GeV)$^{-1}$, as a
function of proton momentum (solid line) in the 22.5 kton mass of SK.
Note that the spectrum falls very quickly (vertical log scale).  The
\v{C}erenkov threshold for protons in water is also shown (thin dotted
vertical line).}
\end{figure} 
%%%%%%%%%%%%%%

The struck proton momentum spectrum in SK for no oscillations (or
oscillations among active flavors only) is
\begin{equation}
\label{dNdp}
\frac{dN}{dp}(p) = 
Z \! \! \int\limits_{(E_{\nu})_{min}}^{\infty} \! \! dE_{\nu}
\; \frac{d N_\nu}{d E_\nu}(E_\nu) \; \frac{d\sigma_{el}}{dp}(E_{\nu},p)
\end{equation}
where $Z = 7.5 \times 10^{33}$ is the number of protons (free and
bound) in the 22.5 kton fiducial mass of SK and $d\sigma_{el}/dp$ is
the differential elastic cross section, Eq.~(\ref{crosssect}).  The
atmospheric neutrino flux~\cite{flux} $d N_\nu/d E_\nu$ has been
integrated over $4 \pi$ (this is done only to calculate the total
yields, as in practice the directionality of the protons can be used).
We sum over all three flavors of neutrinos and antineutrinos, taking
into account the reduced cross section for antineutrinos.  The minimum
neutrino energy for a given proton momentum is
\begin{equation}
\label{Emin}
(E_{\nu})_{min} = \frac{1}{2} (E_p + p - M_p)\,.
\end{equation}
In Fig.~\ref{dNdpI}, we show the complete momentum spectrum for
protons elastically scattered by atmospheric neutrinos and
antineutrinos in SK, per year of detector livetime.

The spectrum falls very steeply, and the fraction of protons above the
\v{C}erenkov threshold at $p = 1.07$ GeV is very small, about 2\%.
For the present exposure time of SK, 1489 days, we predict about 60
protons above the \v{C}erenkov threshold (and about 2000 below).  In
order to normalize our results, we calculated the number of
quasielastic events in SK; we agree with the SK no-oscillation numbers
if we assume a detector efficiency of about $0.7$ (approximately the
official SK number).  Thus in the present data there should be about
40 elastically-scattered protons above the \v{C}erenkov threshold.
This number is small, but it should be noted that much more data is
expected from SK in the future.  And indeed, possibly also from a
future 1 Mton Hyper-Kamiokande detector with $\sim 40$ times higher
rate~\cite{HK} (or UNO, with $\sim 20$ times higher rate~\cite{UNO});
a high-statistics sample of neutrino-proton elastic scattering events
could then be quickly collected.

The small number of protons above the \v{C}erenkov threshold is a
consequence of the large proton mass and the shape of the differential
cross section, which falls steeply above a peak at $p \simeq 400$ MeV,
as shown in Fig.~\ref{dsdp}.  Additionally, the atmospheric neutrino
spectrum is steeply falling with neutrino energy, and this is not
compensated by growth in the cross section, since both the
differential and total elastic cross sections become independent of
neutrino energy above a few GeV.

%%%%%%%%%%%%%%
\begin{figure}
\includegraphics[width=3.25in]{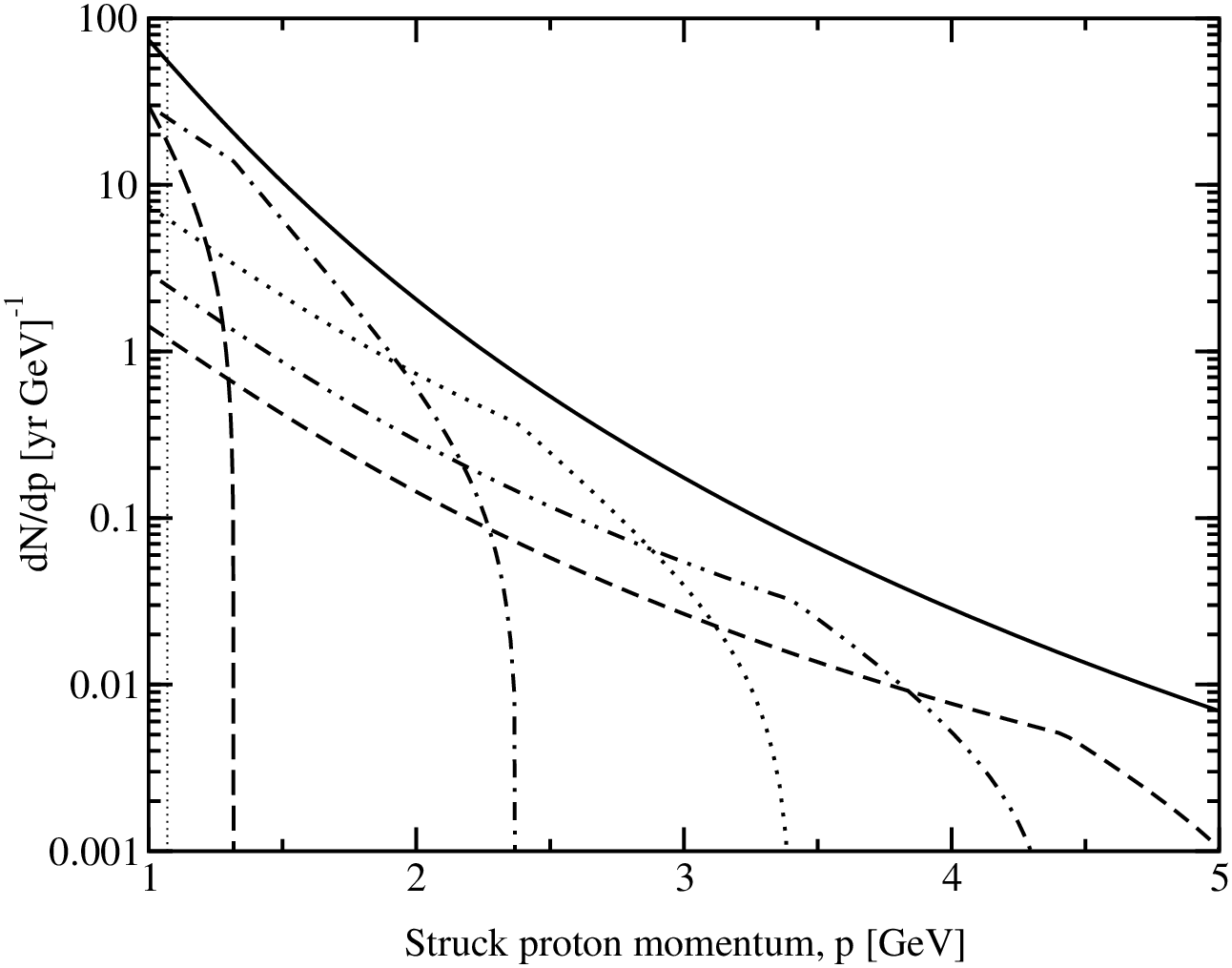}
\caption{\label{dNdpII} Proton spectrum, dN/dp (yr GeV)$^{-1}$, as a
function of proton momentum for different neutrino energy intervals
contributing to it: all neutrinos (solid line), up to 1 GeV (dashed
line), [1, 2] GeV (dot-dashed line), [2, 3] GeV (dotted line), [3, 4]
GeV (doubly-dot-dashed line), [4, 5] GeV (short-dashed line).  Note
that only $p > 1$ GeV is shown.}
\end{figure} 
%%%%%%%%%%%%%%

About 95\% of the protons above threshold are in the interval between
$p = 1.07$ GeV and 2 GeV.  This plays a crucial role in distinguishing
protons from other charged particles, as well as the details of how
they are stopped.  In Fig.~\ref{dNdpII}, we show how different ranges
of neutrino energy contribute to the proton spectrum in this momentum
interval.  About 90\% of the protons above the \v{C}erenkov threshold
are produced by neutrinos with $E_\nu < 5$ GeV, and in fact, the
majority are produced by much lower neutrino energies.  The kinks in
Fig.~\ref{dNdpII} arise because we consider both minimum and a maximum
neutrino energy to draw each curve; e.g., the dotted line for neutrino
energies between 2 and 3 GeV has a kink at $p \simeq 2.4$ GeV because
we do not include neutrino energies below 2 GeV.

%%%%%%%%%%%%%%%%%%%%%%% Neutrino Oscillations %%%%%%%%%%%%%%%%%%%%%%

\subsection{Effect of Neutrino Oscillations}

The vacuum oscillation length is
\begin{equation}
\label{losc}
L_{osc} = \frac{4 \pi E_\nu}{\delta m^2} 
\sim 1000 \left(\frac{E_\nu}{1{\rm\ GeV}}\right) {\rm\ km}
\end{equation}
where we have used $\delta m^2 \simeq 3 \times 10^{-3}$
eV$^2$~\cite{SKatm,SKNC}.  This is close to the distance to the
horizon, so that downgoing neutrinos have not oscillated and upgoing
neutrinos have oscillated several times.  Since the mixing is maximal,
half of the upgoing $\nu_\mu$ remain $\nu_\mu$ and half oscillate to
either $\nu_\tau$ or $\nu_{sterile}$.

As shown in Fig.~\ref{dsdth}, the initial neutrino direction is
largely maintained by the proton direction, the latter to be measured
from its \v{C}erenkov cone.  We show below that the neutrino energy
can also be estimated from the proton \v{C}erenkov information, so
that $L_\nu/E_\nu$ can be estimated on an event-by-event basis, which
improves the ability to study neutrino oscillations.  Even in the
absence of a neutrino energy estimate, Fig.~\ref{dNdpII} shows that
only a narrow range of neutrino energies contributes to the signal
above the \v{C}erenkov threshold but below where the proton spectrum
is greatly diminished.

There are uncertainties in the neutrino-proton elastic scattering
cross section, e.g., from the axial form factor as well as from
nuclear corrections to the free-proton cross section.  There are also
uncertainties introduced by our simple modeling of SK.  For example,
the number of protons above the \v{C}erenkov threshold is quite
sensitive to the index of refraction; we assumed 1.33, but in a more
careful treatment one would have to model the wavelength dependence of
the \v{C}erenkov emission, attenuation, index of refraction, and
phototube quantum efficiency.  Finally, there is also a 20\%
uncertainty on the atmospheric neutrino flux normalization.  In light
of these facts, we must focus on a normalization-independent
observable such the zenith angle spectrum shape, or at least an
up-down ratio.

Consider an initial atmospheric neutrino flavor ratio of
$\nu_e:\nu_\mu:\nu_\tau = 1:2:0$, which is a good approximation.
Downgoing neutrinos have not oscillated, and have these flavor ratios.
However, the upgoing neutrinos have oscillated several times.  For
maximal mixing of $\nu_\mu$ to either $\nu_\tau$ or $\nu_{sterile}$,
the flavor ratios for the upgoing events are either $1:1:1$ or $1:1:0$
(we ignore mixing with $\nu_e$ as well as matter effects).  Since this
is a neutral-current cross section, equally sensitive to all flavors,
the upgoing flux divided by the downgoing flux would be 1 for pure
active oscillations and 2/3 for pure sterile oscillations.  Assuming
40 events in the present SK data, this corresponds to 20 downgoing
events and either 20 (active case) or 13 (sterile case) upgoing
events, the latter reflecting a 1.5 sigma deviation.  Thus with the
present data this technique could not be decisive, but none of the
three techniques used by SK to distinguish $\nu_\tau$ from
$\nu_{sterile}$ is individually decisive~\cite{SKNC}.  The advantage
of neutrino-proton elastic scattering is that it could be rather
clean, both in concept and in practice.

%%%%%%%%%%%%%%%%%%%%%%%%%%%%%%%%%%%%%%%%%%%%%%%%%%%%%%%%%%%%%%%%%%%%%%%%%%%
%                         Experimental PID                                %
%%%%%%%%%%%%%%%%%%%%%%%%%%%%%%%%%%%%%%%%%%%%%%%%%%%%%%%%%%%%%%%%%%%%%%%%%%%

\section{Proton Particle Identification}

%%%%%%%%%%%%%%%%%%%%%%%%%%%% Muons and Electrons %%%%%%%%%%%%%%%%%%%%%%%%%%

\subsection{Electron and Muon PID}

In this section, we show how relativistic protons from neutrino-proton
elastic scattering can be separated from other single-ring events in a
\v{C}erenkov detector like SK.  Quasielastic interactions of
atmospheric neutrinos create relativistic electrons and muons that
produce \v{C}erenkov radiation, which is seen by phototubes as rings
on the walls of the detector.  The rates are large, of order $10^3$
events per year, to be compared to of order 10 relativistic protons
per year.  However, the unique particle identification (PID)
properties of protons will allow rejection of these backgrounds.

Electrons and muons are stopped by continuous electromagnetic energy
losses (mostly ionization, but also radiative losses for electrons);
\v{C}erenkov radiation does not cause significant energy loss.  The
continuous energy loss $-dE/dx$ is given by the Bethe-Bloch equation,
reviewed in Ref.~\cite{RPP}.  The range of a charged particle, the
distance required to bring it to rest, is obtained immediately by
integration.  In Fig.~\ref{RMH2Opm}, we show the range of muons in
water as a function of momentum.  Electrons, because of their small
mass, have higher $-dE/dx$ for the same momentum and are stopped in
less distance; in addition, multiple scattering changes their
direction.  We also show a range curve for protons, which would be
correct if protons only lost energy electromagnetically (at high
momenta, where $\beta = 1$, the proton and muon range are nearly the
same; at low momenta, the proton velocity is less and hence the
electromagnetic losses are higher).  However, for protons in the
relevant momentum range, discrete nuclear collisions are more
important than continuous electromagnetic energy losses.

Since most atmospheric neutrinos are at energies of at least a GeV,
the electrons and muons created in quasielastic reactions almost all
have an initial velocity of $\beta \simeq 1$.  This corresponds to the
maximum \v{C}erenkov intensity, and a \v{C}erenkov emission angle of
$41^\circ$ in water.  For muons, the outer edge of the \v{C}erenkov
rings is sharp, but for electrons, which suffer changes in direction
due to multiple scattering, the outer edge has a more fuzzy
appearance.  As the velocity decreases, the \v{C}erenkov angle and
intensity both decrease.  In an ideal detector, the rings on the wall
would fill in completely as the particle slowed down (and approached
the wall).  In practice, the rapid decrease in the \v{C}erenkov angle
and intensity once the particle falls below $\beta \simeq 1$ mean that
the inner edge of the \v{C}erenkov rings is typically undersampled.
This is especially true for electrons, which lose velocity in less
distance than muons.  Thus muon rings have a sharp outer edge and are
partially filled in, whereas electron rings are fuzzy and not filled
in.  Electrons and muons can be very reliably distinguished in SK (to
about 1\%), as has been confirmed by a variety of means, including
direct beam tests at KEK~\cite{Kasuga}.

%%%%%%%%%%%%%%%%%%%%%%% Proton nuclear collisions %%%%%%%%%%%%%%%%%%%%%%%

\subsection{Proton PID: Nuclear Collisions Only}

Unlike electrons and muons at these momenta, protons have a large
cross section for nuclear collisions.  If electromagnetic energy
losses can be ignored, as we assume in this subsection, then the
fraction of protons surviving a distance $x$ without undergoing a
nuclear collision is
\begin{equation}
\label{exp}
N(x)/N(0) = \exp(-x/\lambda_N)\,.
\end{equation}
For the nuclear attenuation length $\lambda_N$ in water, we use
\begin{equation}
\lambda_N =
\left[
\left(\rho/M_{{\rm H}_2{\rm O}}\right) \; \sigma_{p+{\rm H}_2{\rm O}}
\right]^{-1}\,,
\end{equation}
where $\rho/M_{{\rm H}_2{\rm O}}$ is the number density of water
molecules.  For the cross section, we use
\begin{equation}
\label{csH2O}
\sigma_{p+{\rm H}_2{\rm O}} =
\sigma_{p+^{16}{\rm O}} + 2 \sigma_{p+p_{free}}\,.
\end{equation}
Note that we cannot simply use the number density of nucleons in
water, since nucleons bound in nuclei shadow each other (since nuclear
densities are approximately constant, the cross sectional area of a
nucleus scales as $A^{2/3}$).  For scattering from $^{16}$O, we use
only the reaction (inelastic) cross section, taken from
Ref.~\cite{LAHET} (which corrects the earlier Ref.~\cite{param}).  We
have not included the elastic part of the cross section on $^{16}$O,
as it is very strongly peaked in the forward
direction~\cite{Feshbach}, corresponding to minimal momentum loss (for
a proton with $p = 2$ GeV, a scattering angle less than 20 degrees
corresponds to less than 1\% change in momentum).  For scattering from
free protons, we do use the total cross section, taken from
Ref.~\cite{pp,RPP}; since the target and projectile have the same
mass, it is easier to have substantial momentum transfer.  Above 1 GeV
momentum, the proton nuclear cross section on water is nearly constant
at about 390 mb; the oxygen reaction cross section is about 300 mb,
and the proton cross section is about 45 mb.

\begin{figure}
\includegraphics[width=3.25in]{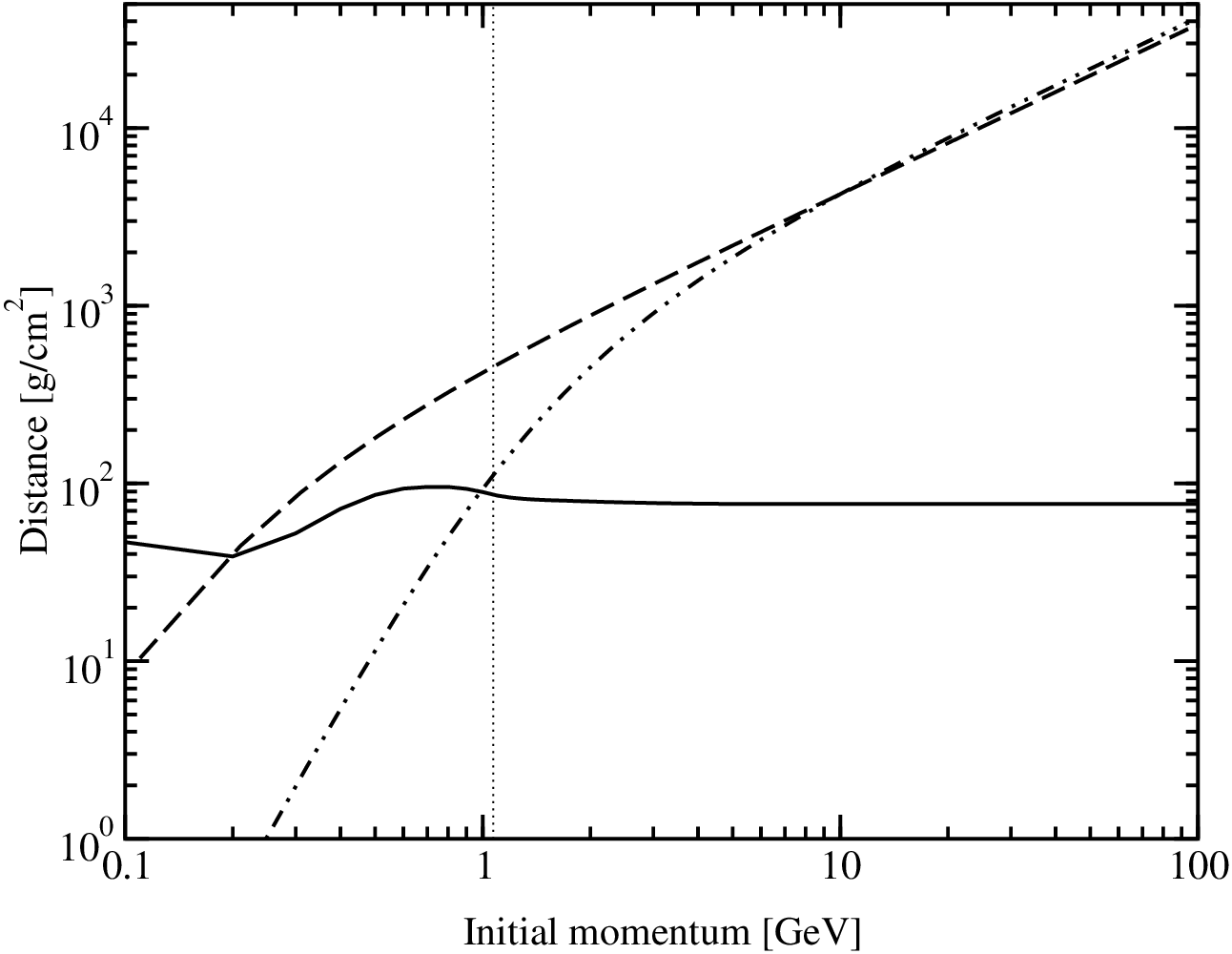}
\caption{\label{RMH2Opm} Distance ($g/cm^2$) traveled in water by
protons and muons as a function of momentum.  We show the range (the
distance to come completely to rest by electromagnetic losses alone)
for muons (dashed line) and protons (dot-dashed line).  For protons,
we also show the nuclear attenuation length $\lambda_N$ in water
(solid line), the distance over which a fraction $1/e$ of protons
travel without scattering.  The \v{C}erenkov threshold for protons is
also shown (thin dotted vertical line).}
\end{figure}

In Fig.~\ref{RMH2Opm}, we show the nuclear attenuation length for
protons in water, ignoring electromagnetic losses.  Above the
\v{C}erenkov threshold, this length is always shorter than the
electromagnetic range of protons calculated if nuclear collisions are
ignored.  Thus in this subsection we consider that only nuclear
collisions are important, and in the next subsection we calculate the
corrections due to electromagnetic energy losses.  In either case,
protons have a short path length and will be fully-contained events.

Since 95\% of the proton events are in the momentum range of $1-2$
GeV, with a steeply falling spectrum (see Fig.~\ref{dNdpII}), a single
nuclear collision with even moderate momentum transfer brings the
proton below the \v{C}erenkov threshold.  In an inelastic collision
with an oxygen nucleus that breaks it into several fragments, it is
very unlikely than any of them are above the \v{C}erenkov threshold.
In a collision with a free proton, the initial momentum of less than 2
GeV is shared between both protons, leaving them both below the
\v{C}erenkov threshold (a forward elastic collision might exchange the
projectile and target protons, causing little change in the
\v{C}erenkov pattern from no collision at all).  Thus we assume that
after a nuclear collision, there are no relativistic protons, neither
the original proton nor any accelerated target protons.

%%%%%%%%%%%%%%
%%%%%%%%%%%%%%

Now we consider the \v{C}erenkov signatures of the struck protons,
assuming that a single nuclear collision brings the proton below the
\v{C}erenkov threshold.  Protons have several unique PID
characteristics.  Since the proton velocity is constant until that
collision, the \v{C}erenkov angle and intensity are constant until
they abruptly vanish.  While the protons are relativistic, they do
have $\beta < 1$, so their \v{C}erenkov angle is less than the
$41^\circ$ for relativistic electrons and muons.  Just as for muons,
the outer edge of the \v{C}erenkov rings is sharp for protons.
However, since their \v{C}erenkov angle is both smaller and constant,
proton rings are filled in very densely and at a constant rate.  The
proton path length is rather short, of order $\lambda_N \simeq 80$ cm,
compared to the several meters typical of muons, and thus the proton
events are always fully contained.  When the proton is abruptly
stopped, the inner edge of the \v{C}erenkov rings is also sharp,
unlike for muons or electrons.

\begin{figure}
\includegraphics[width=3.25in]{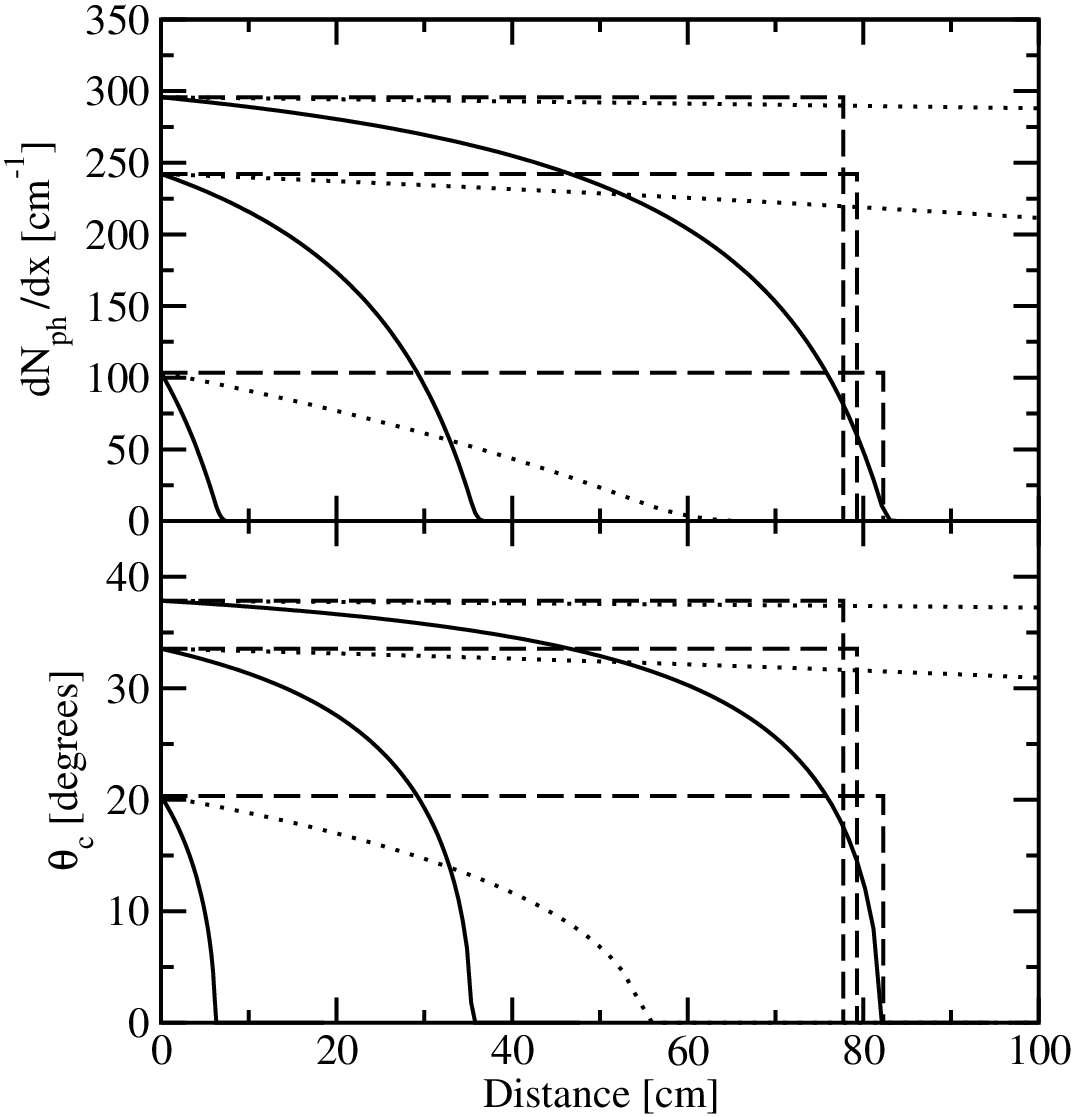}
\caption{\label{dNdxth} \v{C}erenkov intensity (upper plot) and angle
(lower plot) as a function of the traveled path length in water for
muons (solid lines) and protons, considering only nuclear collisions
(dashed lines) or only electromagnetic losses (dotted lines).  From
bottom to top, the initial velocities are $\beta = 0.8$, 0.9, and
0.95.  The \v{C}erenkov intensity is calculated for visible light,
without attenuation or detection efficiency.  With electromagnetic
losses neglected, individual protons always stop abruptly, but with a
distribution of path lengths, Eq.~(\ref{exp}); we used the average
path length $\lambda_N$ in the figure.}
\end{figure}

The number of \v{C}erenkov photons produced per unit path length and
photon wavelength interval by a particle of unit charge and velocity
$\beta$ is
\begin{equation}
\label{dNdxdl}
\frac{d^2N_{ph}}{dx d\lambda} = 
\frac{2 \pi \alpha}{\lambda^2}
\; \left( 1 - \frac{1}{\beta^2 n^2 (\lambda)} \right)
\end{equation}
In Fig.~\ref{dNdxth} we compare how the \v{C}erenkov intensity and
angle vary with the distance traveled for muons (electromagnetic
losses only) and protons (nuclear collisions only).  We choose the
same initial velocities (and hence the same initial \v{C}erenkov angle
and intensity), using $\beta = 0.8$, 0.9, and 0.95, to highlight how
the muon and proton stopping mechanisms differ.  These correspond to
muon momenta of 140 MeV, 220 MeV, and 320 MeV; and to proton momenta
of 1.25 GeV, 1.95 GeV, and 2.85 GeV.  Muons of a given $\beta$ travel
a well-defined distance before they fall below the \v{C}erenkov
threshold.  However, protons of a given $\beta$ travel a variety of
distances, sampled from the distribution in Eq.~(\ref{exp}).  In
Fig.~\ref{dNdxth} we have adopted $\lambda_N$ as the path length for
protons, since this is the average value.

%%%%%%%%%%%%%%
%%%%%%%%%%%%%%

Most muons in the SK atmospheric neutrino data have much longer track
lengths than shown here, and hence are distinguishable.  Those muons
shown in Fig.~\ref{dNdxth} have path lengths short enough to be
confused with protons, but have very different \v{C}erenkov
characteristics.  In addition, excepting the 20\% of negative muons
that capture on oxygen, fully contained muons can be tagged by their
subsequent decay to an electron or positron of up to 53 MeV.  Thus it
should be possible to distinguish protons from muons with very high
efficiency on an event-by-event basis.  In some SK
Ph.D. theses~\cite{SKtheses} the neutrino-proton elastic scattering
cross section was included in the atmospheric neutrino Monte Carlo.
Those events were automatically classified as $e$-like or $\mu$-like
and then considered to be buried by the much larger quasielastic event
samples.  Most were classified as $\mu$-like, which is why we have
emphasized distinguishing protons and muons.  In fact, protons should
be quite distinguishable from electrons as well.

Since the nuclear cross section is nearly constant, the proton track
length is a poor estimator of the proton momentum, and event-by-event
track length fluctuations from Eq.~(\ref{exp}) are more important.
However, the proton momentum can be reliably estimated by the constant
\v{C}erenkov angle and intensity, since those vary appreciably in the
momentum range considered, where $\beta < 1$.

%%%%%%%%%%%%%%%%%%%%%%% And Proton dE/dx %%%%%%%%%%%%%%%%%%%%%%%%%%%%

\subsection{Proton PID: Inclusion of $-dE/dx$ Effects}

From Fig.~\ref{RMH2Opm}, it is clear that for protons with momentum
somewhat above the \v{C}erenkov threshold, electromagnetic losses can
be neglected, as we have assumed.  However, right at the \v{C}erenkov
threshold, this is no longer true, since the $-dE/dx$ losses are
greater and the range is as small as the nuclear interaction length.
Thus the proton behavior is different, as it is continuously slowing
down.  It may finally go below the \v{C}erenkov threshold by either
further electromagnetic losses or a nuclear collision.  In
Fig.~\ref{dNdxth}, we show results for nuclear collisions only (dashed
lines), as well as for electromagnetic losses only (dotted lines).

%%%%%%%%%%%%%%
\begin{figure}
\includegraphics[width=3.25in]{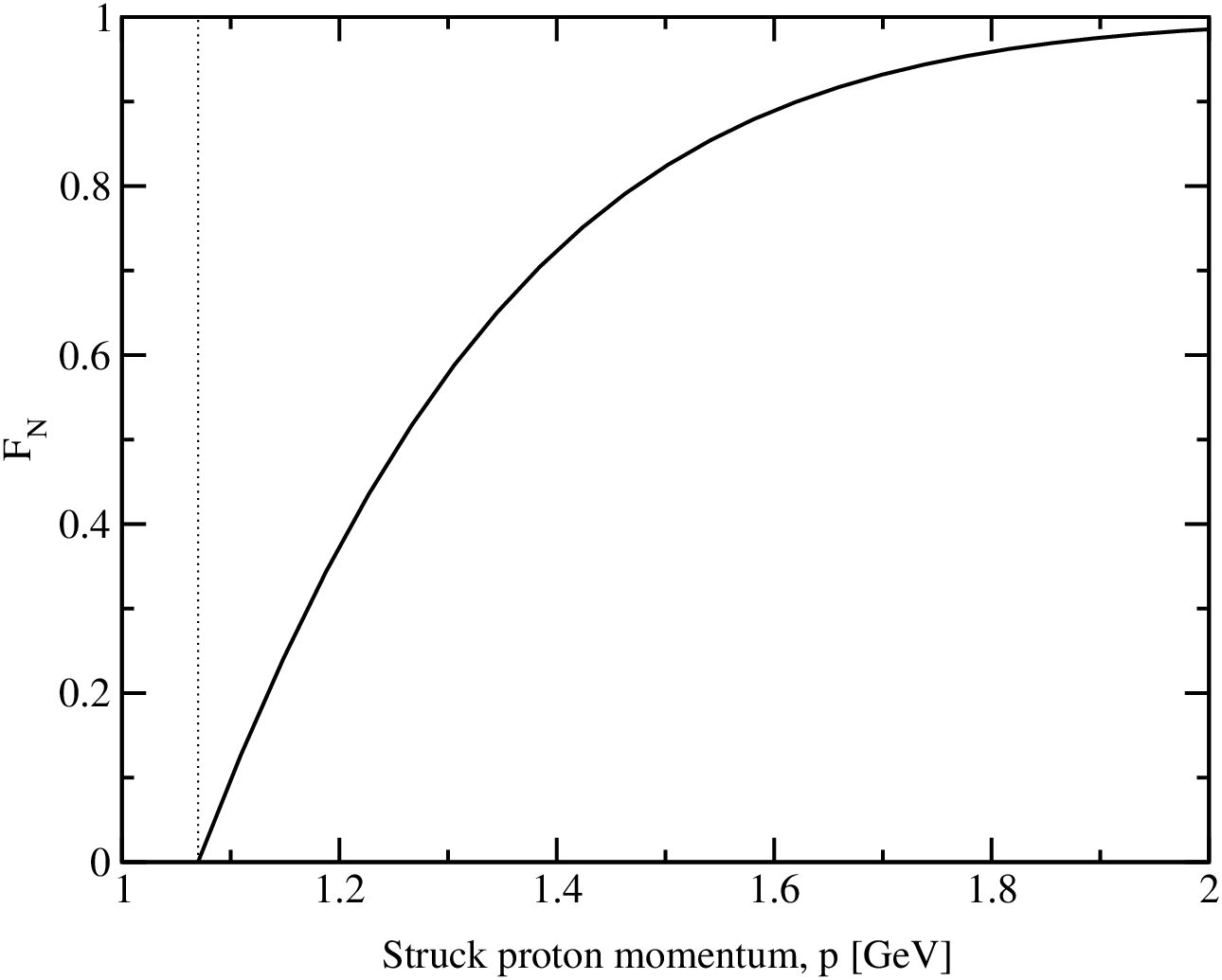}
\caption{\label{frac} The fraction of protons that go below the
\v{C}erenkov threshold by a nuclear collision as a function of proton
momentum.  The \v{C}erenkov threshold for protons in water is also
shown (thin dotted vertical line)}
\end{figure} 
%%%%%%%%%%%%%%

Since the nuclear interaction length is nearly independent of
momentum, the fraction $F_N$ of protons brought below the \v{C}erenkov
threshold by a single nuclear collision is:
\begin{equation}
\label{fraction}
F_N (p) \simeq 1 - e^{-r(p)/\overline{\lambda}_N}
\end{equation}
where $r(p)$ is the distance a proton travels before going below the
\v{C}erenkov threshold, considering only electromagnetic losses, and
$\overline{\lambda}_N = 80$ cm.  Note $r(p)$ is not the full
range, and a 1 GeV proton still travels about a meter, though
invisibly.  The fraction $F_N$ is shown in Fig.~\ref{frac}.
Lower-momentum protons are affected most by electromagnetic energy
losses.  Protons with $p \gtrsim 1.25$ GeV are more likely to be
brought below \v{C}erenkov threshold by a nuclear collision than by
electromagnetic losses.  In fact, if we convolve this curve with the
falling spectrum of struck protons, Fig.~\ref{dNdpI}, about half of
all protons are brought below \v{C}erenkov threshold by a nuclear
collision.

For the protons brought below threshold by a nuclear collision, the
discussion above about proton PID neglecting electromagnetic losses is
the most appropriate, though in some cases there will be some slight
decrease in the \v{C}erenkov angle and intensity before they abruptly
vanish.  The remaining protons fall below the \v{C}erenkov threshold
more gradually, e.g., the lowest momentum case ($\beta = 0.8$) in
Fig.~\ref{dNdxth}, for which electromagnetic losses dominate.
Nevertheless, the \v{C}erenkov behavior is still quite different from
muons.  For muons at the same initial $\beta$, protons go much farther
and produce much more \v{C}erenkov light.  For muons that travel the
same distance above the \v{C}erenkov threshold, the proton
\v{C}erenkov angle is much smaller and falls off more slowly with
distance (the proton velocity is also less, leading to a higher
\v{C}erenkov ring density on the wall).  Thus even low-momentum
protons should be quite distinct from muons (and electrons).  Finally,
when electromagnetic losses dominate, the proton momentum can be
estimated by conventional \v{C}erenkov techniques.

%%%%%%%%%%%%%%%%%%%%%%%% Possible Complications %%%%%%%%%%%%%%%%%%%%%%%

\subsection{Possible Complications}

Since the number of neutrino-proton elastic scattering events above
the \v{C}erenkov threshold is small, of order 10 per year in SK,
careful consideration of backgrounds will be necessary.  Above, we
have motivated the case that protons are distinguishable from the more
numerous electrons and muons from the quasielastic channel; we now
consider other possible backgrounds.

One possible background is from the quasielastic channel, $\nu_\mu + n
\rightarrow \mu^- + p$, but with the muon below the \v{C}erenkov
threshold and the proton above it.  However, the total number of
quasielastic events with a sub-\v{C}erenkov muon is very low, about 35
per year in SK~\cite{relic}.  Most of these are produced by low-energy
neutrinos; for high-energy neutrinos, the fraction with a relativistic
proton is only about 2\%.  Thus this background is negligible.

There are also neutrino-neutron elastic scattering events (in fact,
the cross section is about 1.5 times larger than for protons) in which
the struck neutron carries a large momentum.  Such events are of
course invisible in SK.  However, the struck neutrons can sometimes
scatter a proton with enough momentum transfer that the proton is
above the \v{C}erenkov threshold.  In the E734 accelerator neutrino
experiment, it was estimated that such events were about 15\% of the
measured neutrino-proton elastic scattering signal.  It should be less
here since that tracking calorimeter had a lower (sub-\v{C}erenkov)
threshold for protons.  Note also that such events would partially
compensate the loss of neutrino-proton elastic scattering events from
nuclear reinteractions.

We ignore the production of pions, in the initial interaction, by
proton reinteraction in the initial nucleus, or in the final nuclear
collision.  Pions produced in the initial interaction are not part of
the neutrino-proton elastic scattering channel, and cause multi-ring
events.  We are only considering single-ring events.  Neutral-current
single-pion events with the pion absorbed in the nucleus could be a
background to the elastic channel; however, the fraction of these
events with a proton above the \v{C}erenkov threshold should be even
lower than 2\%, due to kinematics.  Monte Carlo and real data on on
quasielastic scattering using accelerator neutrinos in the K2K SciFi
detector suggest that secondary pion contributions are
minimal~\cite{Walter}.  In their Monte Carlo results, which have a
much more complete treatment of the physics than we have presented
here, the track multiplicity was always 1 (muon) or 2 (muon and
proton), and never 3 (including secondary pions).  Additionally, pions
created in the final nuclear collision would be delayed from the
initial proton by several nanoseconds.  The figures in
Ref.~\cite{Walter} also support our assumption that when a
high-momentum proton has a nuclear collision, it suffers a large
momentum loss without accelerating new protons.

Another possible source of background is atmospheric muons interacting
with the surrounding rock and producing fast neutrons that can enter
the detector without triggering the veto.  These neutrons could in
principle scatter protons above the \v{C}erenkov threshold.  Most
neutrons are far too low in energy to be effective~\cite{wolf}
and neutrons are strongly attenuated by the 4.5 m water shielding.
Incidentally, neutrino-proton elastic scattering events might be visible
in the Soudan-2 experiment~\cite{Soudan2}, which has much less mass
and shielding than SK but can detect lower-energy (sub-\v{C}erenkov)
protons in a tracking calorimeter.  The number of events could be a
few tens, but the neutron backgrounds could be
comparable~\cite{sterilePi}.  We are not aware of any official
analysis of these events by the Soudan-2 collaboration.  Fast neutrons
from the walls can also produce neutral pions by nuclear collisions in
the detector; if the two photon rings from the decay are overlapping,
this can resemble an atmospheric $\nu_e$ event~\cite{Olga}.  However,
the SK collaboration has shown that such events contribute less than
0.1\% of the atmospheric $\nu_e$ signal~\cite{neutrons,Kajita}.

In summary, a full Monte Carlo study will be needed to correctly
implement the initial neutrino interactions, possible nuclear
reinteractions, pions, nuclear stopping and electromagnetic losses,
backgrounds, and most importantly the PID in a realistic detector.
Nevertheless, we believe that it looks promising that the relatively
few (of order 10 per year in SK) neutrino-proton elastic scattering
events above the \v{C}erenkov threshold can be detected with little
background.

%%%%%%%%%%%%%%%%%%%%%%%%%%%%%%%%%%%%%%%%%%%%%%%%%%%%%%%%%%%%%%%%%%%%%%%%%
%                        Related Applications                           %
%%%%%%%%%%%%%%%%%%%%%%%%%%%%%%%%%%%%%%%%%%%%%%%%%%%%%%%%%%%%%%%%%%%%%%%%%

\section{Related Applications}

%%%%%%%%%%%%%%%%%%%%%%%%%%%%% Near Detectors %%%%%%%%%%%%%%%%%%%%%%%%%%%%

\subsection{Accelerator Neutrinos: NC and CC Channels}

We are not aware of any experiments with $p = 1-2$ GeV proton beams in
\v{C}erenkov detectors that would test the PID techniques introduced
above.  However, it should be possible to use accelerator neutrino
beams to initiate neutrino-proton elastic scattering events in the
right momentum range.  The spectrum of accelerator neutrinos does not
extend as high in energy as for atmospheric neutrinos (though note
Fig.~\ref{dNdpII} shows that most of the signal comes from low energy
neutrinos), but the total numbers of events expected are much larger.

The K2K 1-kton near detector would be a good place to start, as this
detector is designed to mimic SK~\cite{K2K}.  This data could be very
useful for developing proton PID techniques.  It would also be useful
to study quasielastic events in which the proton is above the
\v{C}erenkov threshold; these are about 8 times more numerous than the
comparable elastic events (though $\sigma_{NC}/\sigma_{CC} \simeq
0.2$, the ratio of the differential cross sections for $p \simeq 1 -
2$ GeV is smaller).  Measuring both the outgoing lepton and proton
would allow reconstruction of the neutrino energy, useful for
measuring the neutrino spectrum.  With $\sim 10^5$ events expected, we
estimate $\sim 10^3$ quasielastic and $\sim 10^2$ elastic events with
a relativistic proton.

The MiniBooNE detector~\cite{MiniBooNE} could also be used.  Since it
is designed to test the LSND signal (small mixing angle and large
$\delta m^2$)~\cite{lsnd}, it can be considered a near detector for
oscillations with the atmospheric $\delta m^2$.  It contains 0.68
ktons of mineral oil, and is primarily a \v{C}erenkov detector (about
3:1 \v{C}erenkov to scintillation light).  MiniBooNE has unique
characteristics that will help proton PID.  The index of refraction in
oil ($\simeq 1.5$) is larger than in water, allowing a lower
\v{C}erenkov threshold (and larger angle and intensity).  The density
of oil ($\simeq 0.8$ g/cm$^3$) is less than for water, which means
longer tracks.  And once a proton falls below the \v{C}erenkov
threshold, it still produces scintillation light.  With $\sim 5 \times
10^5$ events expected, we estimate $\sim 5 \times 10^3$ quasielastic
and $\sim 5 \times 10^2$ elastic events with a relativistic proton.
These studies would be an appealing complement to plans to measure the
elastic scattering cross section at $Q^2 = 0$, a test of the strange
quark contribution $\Delta s$ to the proton spin~\cite{Tayloe}.  The
combined elastic and quasielastic data could measure the
$Q^2$-dependence of the uncertain axial form factor.

%%%%%%%%%%%%%%%%%%%%%%%%%%%%%% ATM CC %%%%%%%%%%%%%%%%%%%%%%%%%%%%%%%%%%

\subsection{Atmospheric Neutrinos: CC Channel}

As noted, we expect about 8 times more quasielastic than elastic
events with a proton above the \v{C}erenkov threshold.  Thus we
estimate about $\sim 300$ such events in the 1489-day SK data.  Taking
$\nu_\mu$ oscillations into account would reduce this number since
$\nu_\mu$ oscillate to either $\nu_\tau$ (mostly below the CC
threshold) or $\nu_{sterile}$.  This has a very important consequence
from the point of view of atmospheric neutrino oscillations, as it
would allow the determination of the neutrino energy and direction on
an event-by-event basis, allowing a better measurement of the $L/E$
dependence of oscillations.  Note also that these quasielastic events
with a proton are produced only by neutrinos, and not antineutrinos,
useful to studying matter effects and the neutrino/antineutrino
ratio~\cite{mantle}.

%%%%%%%%%%%%%%%%%%%%%%%%%%%%%%%%%%%%%%%%%%%%%%%%%%%%%%%%%%%%%%%%%%%%%%%%%%
%                             Conclusions                                %  
%%%%%%%%%%%%%%%%%%%%%%%%%%%%%%%%%%%%%%%%%%%%%%%%%%%%%%%%%%%%%%%%%%%%%%%%%%

\section{Conclusions}

We propose that neutrino-proton elastic scattering, $\nu + p
\rightarrow \nu + p$, could be a useful detection reaction for
atmospheric neutrinos in SK.  The fraction of protons above the
\v{C}erenkov threshold is not large, only about 2\%, but there should
be about 40 identifiable events in the present 1489-day data.  We have
shown that it should be possible to separate protons from electrons
and muons, since the relevant protons are not fully relativistic and
will typically be stopped by a single nuclear collision.  Proton
\v{C}erenkov rings have sharp outer and inner edges, are very densely
filled, and correspond to a short path length and small \v{C}erenkov
angle.  These are fully-contained, single-ring events.  In order to
test our proposal, a detailed detector Monte Carlo simulation will be
needed.

Neutrino-proton elastic scattering is a neutral-current reaction and
so measures the total active neutrino flux.  For the relevant neutrino
energies, oscillations occur at the distance to the horizon.  In
addition, protons above the \v{C}erenkov threshold preserve the
neutrino direction.  These facts mean that this data can be used to
test atmospheric $\nu_\mu \rightarrow \nu_\tau$ versus $\nu_\mu
\rightarrow \nu_{sterile}$ oscillations.  Since there are
normalization uncertainties in the atmospheric neutrino flux, the
cross section, and aspects of the detection, an up-down asymmetry test
should be used.  Let us assume 40 identifiable events in the present
SK data (no oscillations).  With oscillations, there should be 20
downgoing events and either 20 ($\nu_\mu \rightarrow \nu_\tau$) or 13
($\nu_\mu \rightarrow \nu_{sterile}$) upgoing events.  While not
decisive, other techniques for active-sterile discrimination are not
individually decisive either; they obtain their power in combination.
Neutrino-proton elastic scattering has the advantage of being clean in
concept.  The rate in the proposed Hyper-Kamiokande detector would be
about 40 times larger~\cite{HK}.

Our results on neutrino-proton elastic scattering have other immediate
and important applications.  First, using accelerator neutrinos, this
channel can be seen in the K2K 1-kton near detector and in MiniBooNE.
This data will reduce the cross section uncertainties and develop the
proton PID techniques.  Tagging relativistic protons will be similarly
useful in the quasielastic channel in these detectors.  Second, for a
small fraction of the atmospheric neutrino quasielastic events, the
proton is relativistic and can be tagged using the techniques
presented here.  This uniquely selects $\nu_\mu$ (not $\bar{\nu}_\mu$)
events, useful for understanding matter effects, and allows
determination of the neutrino energy and direction, useful for
studying the $L/E$ dependence of oscillations.

%%%%%%%%%%%%%%%%%%%%%%%%%%%%%%%%%%%%%%%%%%%%%%%%%%%%%%%%%%%%%%%%%%%%%%%%%
%                            Acknowledgments                            %
%%%%%%%%%%%%%%%%%%%%%%%%%%%%%%%%%%%%%%%%%%%%%%%%%%%%%%%%%%%%%%%%%%%%%%%%%

\section*{ACKNOWLEDGMENTS}

We thank Steve Brice, Takaaki Kajita, Paolo Lipari, Mark Messier,
Georg Raffelt, Michael Smy, Manuel Vicente-Vacas, and Sam Zeller for
useful discussions.  We are especially grateful to Mark Vagins for
extensive discussions.  JFB is supported as the David N. Schramm
Fellow by Fermilab (operated by URA under DOE contract
No. DE-AC02-76CH03000), and additionally by NASA under NAG5-10842.
SPR is supported by the Spanish MCED with a graduate FPU fellowship,
by the Spanish MCYT and FEDER European Funds through the project
FPA2002-00612, and by the OCYT of the Generalitat Valenciana under the
Grant GV01-94.  SPR thanks the Fermilab Theoretical Astrophysics Group
for hospitality.

%%%%%%%%%%%%%%%%%%%%%%%%%%%%%%%%%%%%%%%%%%%%%%%%%%%%%%%%%%%%%%%%%%%%%%%%
%                              Bibliography                            %
%%%%%%%%%%%%%%%%%%%%%%%%%%%%%%%%%%%%%%%%%%%%%%%%%%%%%%%%%%%%%%%%%%%%%%%%


\begin{thebibliography}{100}

\bibitem{SKatm}
Y.~Fukuda {\it et al.},
%``Measurement of a small atmospheric nu/mu / nu/e ratio,''
Phys.\ Lett.\ B {\bf 433}, 9 (1998);
%%CITATION = HEP-EX 9803006;%%
Y.~Fukuda {\it et al.},
%``Study of the atmospheric neutrino flux in the multi-GeV energy range,''
Phys.\ Lett.\ B {\bf 436}, 33 (1998);
%%CITATION = HEP-EX 9805006;%%
Y.~Fukuda {\it et al.},
%``Neutrino-induced upward stopping muons in Super-Kamiokande,''
Phys.\ Lett.\ B {\bf 467}, 185 (1999);
%%CITATION = HEP-EX 9908049;%%
Y.~Fukuda {\it et al.},
%``Measurement of the flux and zenith-angle distribution of upward
%through-going muons by Super-Kamiokande,'' 
Phys.\ Rev.\ Lett.\  {\bf 82}, 2644 (1999).
%%CITATION = HEP-EX 9812014;%%

\bibitem{reactor}
M.~Apollonio {\it et al.},
%``Limits on neutrino oscillations from the CHOOZ experiment,''
Phys.\ Lett.\ B {\bf 466}, 415 (1999);
%%CITATION = HEP-EX 9907037;%%
F.~Boehm {\it et al.},
%``Final results from the Palo Verde neutrino oscillation experiment,''
Phys.\ Rev.\ D {\bf 64}, 112001 (2001).
%%CITATION = HEP-EX 0107009;%%

\bibitem{fournulab}
M.~Maltoni, T.~Schwetz, M.~A.~Tortola and J.~W.~Valle,
%``Ruling out four-neutrino oscillation interpretations of the LSND anomaly?,''
Nucl.\ Phys.\ B {\bf 643}, 321 (2002);
%%CITATION = HEP-PH 0207157;%%
H.~Pas, L.~G.~Song and T.~J.~Weiler,
%``The hidden sterile neutrino and the (2+2) sum rule,''
hep-ph/0209373;
%%CITATION = HEP-PH 0209373;%%
and references therein.

\bibitem{lsnd}
A.~Aguilar {\it et al.}, Phys.\ Rev.\ D {\bf 64}, 112007 (2001).
%%CITATION = HEP-EX 0104049;%%

\bibitem{fournucosmo}
P.~Di Bari,
%``Update on neutrino mixing in the early universe,''
Phys.\ Rev.\ D {\bf 65}, 043509 (2002);
%%CITATION = HEP-PH 0108182;%%
K.~N.~Abazajian,
%``Telling three from four neutrinos with cosmology,''
astro-ph/0205238;
%%CITATION = ASTRO-PH 0205238;%%
and references therein.

\bibitem{sterilePi}
F.~Vissani and A.~Y.~Smirnov,
%``Neutral-to-charged current events ratio in atmospheric neutrinos and
%neutrino oscillations,'' 
Phys.\ Lett.\ B {\bf 432}, 376 (1998).
%%CITATION = HEP-PH 9710565;%%

\bibitem{sterileME}
Q.~Y.~Liu and A.~Y.~Smirnov,
%``Neutrino mass spectrum with nu/mu $\to$ nu/s oscillations of
%atmospheric neutrinos,''
Nucl.\ Phys.\ B {\bf 524}, 505 (1998);
%%CITATION = HEP-PH 9712493;%%
R.~Foot, R.~R.~Volkas and O.~Yasuda,
%``Comparing and contrasting the nu/mu $\to$ nu/tau and nu/mu $\to$ nu/s
% solutions to the atmospheric neutrino problem with SuperKamiokande data,''
Phys.\ Rev.\ D {\bf 58}, 013006 (1998);
%%CITATION = HEP-PH 9801431;%%
P.~Lipari and M.~Lusignoli,
%``Comparison of nu/mu <--> nu/tau and nu/mu <--> nu/s oscillations as  
%solutions of the atmospheric neutrino problem,''
Phys.\ Rev.\ D {\bf 58}, 073005 (1998).
%%CITATION = HEP-PH 9803440;%%
M.~C.~Gonzalez-Garcia, H.~Nunokawa, O.~L.~Peres and J.~W.~Valle,
%``Active-active and active-sterile neutrino oscillation solutions to the  
%atmospheric neutrino anomaly,''
Nucl.\ Phys.\ B {\bf 543}, 3 (1999);
%%CITATION = HEP-PH 9807305;%%
G.~L.~Fogli, E.~Lisi and A.~Marrone,
%``Four-neutrino oscillation solutions of the atmospheric neutrino anomaly,''
Phys.\ Rev.\ D {\bf 63}, 053008 (2001).
%%CITATION = HEP-PH 0009299;%%

\bibitem{sterileMR}
L.~J.~Hall and H.~Murayama,
%``Study of inclusive multi-ring events from atmospheric neutrinos,''
Phys.\ Lett.\ B {\bf 436}, 323 (1998).
%%CITATION = HEP-PH 9806218;%%

\bibitem{sterileTau}
L.~J.~Hall and H.~Murayama,
%``Tau appearance in atmospheric neutrino interactions,''
Phys.\ Lett.\ B {\bf 463}, 241 (1999);
%%CITATION = HEP-PH 9810468;%%
T.~Stanev,
%``Possible tau appearance experiment with atmospheric neutrinos,''
Phys.\ Rev.\ Lett.\  {\bf 83}, 5427 (1999).
%%CITATION = ASTRO-PH 9907018;%%

\bibitem{sterileDK}
S.~Nussinov and R.~Shrock,
%``Weak and electromagnetic nuclear decay signatures for neutrino reactions 
%in Superkamiokande,''
Phys.\ Rev.\ Lett.\  {\bf 86}, 2223 (2001);
%%CITATION = HEP-PH 0009334;%%
E.~Kolbe, K.~Langanke and P.~Vogel,
%``Estimates of weak and electromagnetic nuclear decay signatures for 
%neutrino reactions in Super-Kamiokande,''
Phys.\ Rev.\ D {\bf 66}, 013007 (2002).
%%CITATION = PHRVA,D66,013007;%%

\bibitem{SKNC}
S.~Fukuda {\it et al.},
%``Tau neutrinos favored over sterile neutrinos in atmospheric muon  neutrino
%oscillations,'' 
Phys.\ Rev.\ Lett.\  {\bf 85}, 3999 (2000);
%%CITATION = HEP-EX 0009001;%%
A.~Habig,
%``Discriminating between nu/mu <--> nu/tau and nu/mu <--> nu(sterile) in
%atmospheric nu/mu oscillations with the Super-Kamiokande detector,'' 
hep-ex/0106025.
%%CITATION = HEP-EX 0106025;%%

\bibitem{elastic}
J.~F.~Beacom, W.~M.~Farr and P.~Vogel,
%``Detection of supernova neutrinos by neutrino proton elastic scattering,''
Phys.\ Rev.\ D {\bf 66}, 033001 (2002).
%%CITATION = HEP-PH 0205220;%%

\bibitem{Ahrens}
L.~A.~Ahrens {\it et al.},
%``Precise Determination Of Sin**2-Theta-W From Measurements Of The
%Differential Cross Sections For Muon-Neutrino P $\to$ Muon-Neutrino P And
%Anti-Muon-Neutrino P $\to$ Anti-Muon-Neutrino P,''  
Phys.\ Rev.\ Lett.\  {\bf 56}, 1107 (1986)
[Erratum-ibid.\  {\bf 56}, 1883 (1986)];
%%CITATION = PRLTA,56,1107;%%
L.~A.~Ahrens {\it et al.},
%``Measurement Of Neutrino - Proton And Anti-Neutrino - Proton Elastic
%Scattering,'' 
Phys.\ Rev.\ D {\bf 35}, 785 (1987);
%%CITATION = PHRVA,D35,785;%%
L.~A.~Ahrens {\it et al.},
%``A Study Of The Axial Vector Form-Factor And Second Class Currents In
%Anti-Neutrino Quasielastic Scattering,'' 
Phys.\ Lett.\ B {\bf 202}, 284 (1988).
%%CITATION = PHLTA,B202,284;%%

\bibitem{SKtheses}
M.~D.~Messier,
%``Evidence For Neutrino Mass From Observations Of Atmospheric Neutrinos 
%With Super-Kamiokande,''
Ph.~D. Thesis,  UMI-99-23965;\\
Jun Kameda,
%''Detailed Studies of neutrino oscillation with atmospheric neutrinos of
%wide energy range from 100 MeV to 1000 GeV in Super-Kamiokande''
Ph.~D. Thesis (2002);\\
\url{http://www-sk.icrr.u-tokyo.ac.jp/sk/pub/}.

\bibitem{cs}
C.~H.~Llewellyn Smith,
%``Neutrino Reactions At Accelerator Energies,''
Phys.\ Rept.\  {\bf 3}, 261 (1972).
%%CITATION = PRPLC,3,261;%%

\bibitem{axial2}
A.~Liesenfeld {\it et al.},
%``A measurement of the axial form-factor of the nucleon by the
% p(e,e' pi+)n reaction at W = 1125-MeV,''
Phys.\ Lett.\ B {\bf 468}, 20 (1999).
%%CITATION = NUCL-EX 9911003;%%

\bibitem{axial1}
V.~Bernard, L.~Elouadrhiri and U.~G.~Meissner,
%``Axial structure of the nucleon,''
J.\ Phys.\ G {\bf 28}, R1 (2002).
%%CITATION = HEP-PH 0107088;%%

\bibitem{strange}
G.~T.~Garvey, W.~C.~Louis and D.~H.~White,
%``Determination of proton strange form-factors from neutrino p elastic
%scattering,'' 
Phys.\ Rev.\ C {\bf 48}, 761 (1993);
%%CITATION = PHRVA,C48,761;%%
W.~M.~Alberico, S.~M.~Bilenky and C.~Maieron,
%``Strangeness in the nucleon: Neutrino nucleon and polarized electron 
% nucleon scattering,''
Phys.\ Rept.\  {\bf 358}, 227 (2002);
%%CITATION = HEP-PH 0102269;%%
R.~D.~McKeown and M.~J.~Ramsey-Musolf,
%``The nucleon's mirror image: Revealing the strange and unexpected,''
hep-ph/0203011.
%%CITATION = HEP-PH 0203011;%%

\bibitem{Gaisser}
T.~K.~Gaisser and J.~S.~O'Connell,
%``Interactions Of Atmospheric Neutrinos In Nuclei At Low-Energy,''
Phys.\ Rev.\ D {\bf 34}, 822 (1986).
%%CITATION = PHRVA,D34,822;%%

\bibitem{Co}
C.~Bleve, G.~Co, I.~De Mitri, P.~Bernardini, G.~Mancarella, D.~Martello 
and A.~Surdo,
%``Effects of nuclear re-interactions in quasi-elastic neutrino nucleus  
%scattering,''
Astropart.\ Phys.\  {\bf 16}, 145 (2001);
%%CITATION = NUCL-TH 0012015;%%
G.~Co', C.~Bleve, I.~De Mitri and D.~Martello,
%``Nuclear re-interaction effects in quasi-elastic neutrino nucleus
% scattering,''
Nucl.\ Phys.\ Proc.\ Suppl.\  {\bf 112}, 210 (2002).
%%CITATION = NUCL-TH 0203025;%%

\bibitem{Walter}
C.~W.~Walter,
%``Quasi-elastic events and nuclear effects with the K2K Sci-Fi detector,''
Nucl.\ Phys.\ Proc.\ Suppl.\  {\bf 112}, 140 (2002),
\url{http://neutrino.kek.jp/nuint01/}.
%%CITATION = NUPHZ,112,140;%%

\bibitem{nucodes}
D.~Casper,
%``The nuance neutrino physics simulation, and the future,''
Nucl.\ Phys.\ Proc.\ Suppl.\  {\bf 112}, 161 (2002);
%%CITATION = HEP-PH 0208030;%%
Y.~Hayato,
%``Neut,''
Nucl.\ Phys.\ Proc.\ Suppl.\  {\bf 112}, 171 (2002);
%%CITATION = NUPHZ,112,171;%%
H.~Gallagher,
%``The neugen neutrino event generator,''
Nucl.\ Phys.\ Proc.\ Suppl.\  {\bf 112}, 188 (2002);
%%CITATION = NUPHZ,112,188;%%
\url{http://neutrino.kek.jp/nuint01/}.

\bibitem{flux}
M.~Honda, T.~Kajita, K.~Kasahara and S.~Midorikawa,
%``Calculation of the flux of atmospheric neutrinos,''
Phys.\ Rev.\ D {\bf 52}, 4985 (1995).
%%CITATION = HEP-PH 9503439;%%

\bibitem{HK}
Y.~Itow {\it et al.},
%``The JHF-Kamioka neutrino project,''
hep-ex/0106019.
%%CITATION = HEP-EX 0106019;%%

\bibitem{UNO}
C.~K.~Jung,
%``Feasibility of a next generation underground water Cherenkov
%detector: UNO,''
hep-ex/0005046.
%%CITATION = HEP-EX 0005046;%%

\bibitem{RPP}
K.~Hagiwara {\it et al.},
%``Review Of Particle Physics,''
Phys.\ Rev.\ D {\bf 66}, 010001 (2002).
%%CITATION = PHRVA,D66,010001;%%

\bibitem{Kasuga}
S.~Kasuga {\it et al.},
%``A Study on the e / mu identification capability of a water Cerenkov
%detector and the atmospheric neutrino problem,''
Phys.\ Lett.\ B {\bf 374}, 238 (1996).
%%CITATION = PHLTA,B374,238;%%

\bibitem{LAHET}
R.~E.~Prael and M.~B.~Chadwick,
%``Addendum to: Applications of the evaluated nuclear data in the
% LAHET code,'' 
Los Alamos National Laboratory Report LA-UR-97-1745 (May 6, 1997), 
\url{http://www-xdiv.lanl.gov/XCI/PEOPLE/rep/plist.html}.

\bibitem{param}
H.~P.~Wellisch and D.~Axen,
%``Total Reaction Cross Section Calculations In Proton Nucleus Scattering,''
Phys.\ Rev.\ C {\bf 54}, 1329 (1996).
%%CITATION = PHRVA,C54,1329;%%

\bibitem{Feshbach}
D.~F. Measday and C. Richard-Serre, CERN-69-17;
H.~Feshbach, {\it Theoretical Nuclear Physics: Nuclear Reactions}
(John Wiley \& Sons, Inc., New York, 1992).

\bibitem{pp}
R.~K.~Tripathi, J.~W.~Wilson and F.~A.~Cucinotta,
%``Medium Modified Nucleon Nucleon Cross Sections In A Nucleus,''
Nucl.\ Instrum.\ Meth.\ B {\bf 152}, 425 (1999).
%%CITATION = NUIMA,B152,425;%%

\bibitem{relic}
M.~Malek {\it et al.},
%``Search for supernova relic neutrinos at Super-Kamiokande,''
hep-ex/0209028.
%%CITATION = HEP-EX 0209028;%%

\bibitem{wolf}
F.~Boehm {\it et al.},
%``Neutron production by cosmic-ray muons at shallow depth,''
Phys.\ Rev.\ D {\bf 62}, 092005 (2000);
%%CITATION = HEP-EX 0006014;%%
J.~Wolf,
%``Measurement of muon induced neutron background at shallow sites,''
hep-ex/0211032.
%%CITATION = HEP-EX 0211032;%%

\bibitem{Soudan2}
W.~W.~Allison {\it et al.},
%``The atmospheric neutrino flavor ratio from a 3.9 fiducial kiloton-year  
%exposure of Soudan 2,''
Phys.\ Lett.\ B {\bf 449}, 137 (1999).
%%CITATION = HEP-EX 9901024;%%

\bibitem{Olga}
O.~G.~Ryazhskaya,
%``Is there an excess of electron-neutrinos in the atmospheric flux?,''
JETP Lett.\  {\bf 60}, 617 (1994);
%%CITATION = JTPLA,60,617;%%
O.~G.~Ryazhskaya,
%``Comment On An Interpretation Of Measurements Of The Flux Of Atmospheric 
%Neutrinos By The Kamiokande Detector,''
JETP Lett.\  {\bf 61}, 237 (1995).
%%CITATION = JTPLA,61,237;%%

\bibitem{neutrons}
Y.~Fukuda {\it et al.},
%``Study of neutron background in the atmospheric neutrino sample in
% Kamiokande,''
Phys.\ Lett.\ B {\bf 388}, 397 (1996).
%%CITATION = PHLTA,B388,397;%%

\bibitem{Kajita}
T.~Kajita and Y.~Totsuka,
%``Observation of atmospheric neutrinos,''
Rev.\ Mod.\ Phys.\  {\bf 73}, 85 (2001).
%%CITATION = RMPHA,73,85;%%

\bibitem{K2K}
M.~H.~Ahn {\it et al.},
%``Indications of neutrino oscillation in a 250-km long-baseline experiment,''
hep-ex/0212007;
%%CITATION = HEP-EX 0212007;%%
Y.~Oyama,
%``Present status of the K2K experiment,''
hep-ex/0210030.
%%CITATION = HEP-EX 0210030;%%

\bibitem{MiniBooNE} 
E.~Church {\it et al.}, ``A proposal for an experiment to measure
$\nu_\mu \rightarrow \nu_e$  oscillations and $\nu_\mu$ disappearance at the
Fermilab Booster: BooNE,'' FERMILAB-P-0898; 
I. Stancu {\it et al.}, ``The MiniBooNE Detector Technical Design Report'';
see \url{http://www-boone.fnal.gov/}.

\bibitem{Tayloe}
R.~Tayloe,
%``A Measurement Of The Strange Quark Contribution To The Nucleon Spin 
%Via Neutrino Nucleon Elastic Scattering,''
Nucl.\ Phys.\ Proc.\ Suppl.\  {\bf 105}, 62 (2002).
%%CITATION = NUPHZ,105,62;%%

\bibitem{mantle}
M.~C.~Ba\~nuls, G.~Barenboim and J.~Bernab\'eu,
%``Medium effects for terrestrial and atmospheric neutrino oscillations,''
Phys.\ Lett.\ B {\bf 513}, 391 (2001);
%%CITATION = HEP-PH 0102184;%%
J.~Bernab\'eu, S.~Palomares-Ruiz, A.~P\'erez and S.~T.~Petcov,
%``The earth mantle-core effect in matter-induced asymmetries for
% atmospheric neutrino oscillations,'' 
Phys.\ Lett.\ B {\bf 531}, 90 (2002);
%%CITATION = HEP-PH 0110071;%%
T.~K.~Gaisser and T.~Stanev,
%``Charge ratio of muons from atmospheric neutrinos,''
astro-ph/0210512.
%%CITATION = ASTRO-PH 0210512;%%

\end{thebibliography}
\end{document}